 \documentstyle[12pt]{article}
  
\catcode`@=11
\def\secteqno{\@addtoreset{equation}{section}%
\def\theequation{\thesection.\arabic{equation}}}
\catcode`@=12
\secteqno
\newcommand{\be}{\begin{equation}}
\newcommand{\ee}{\end{equation}}
\newcommand{\bea}{\begin{eqnarray}}
\newcommand{\eea}{\end{eqnarray}}
\newcommand{\bref}[1]{(\ref{#1})}
\newcommand{\nn}{\nonumber}

\begin{document}

\begin{flushright}
\parbox{4.2cm}
{2013,~May 28\\
KEK-TH-1629 \hfill \\
}
\end{flushright}

\vspace*{1.1cm}

\begin{center}
 \Large\bf   M5 algebra and SO(5,5) duality  
\end{center}
\vspace*{1.5cm}
\centerline{\large Machiko Hatsuda$^{\dagger\ast a}$ and Kiyoshi Kamimura$^{\star b}$}
\begin{center}
$^{\dagger}$\emph{Physics Department, Juntendo University, 270-1695, Japan}
\\
$^{\ast}$\emph{KEK Theory Center, High Energy Accelerator Research 
Organization,\\
Tsukuba, Ibaraki 305-0801, Japan} 
\\
$^{\star}$\emph{Department of Physics, Toho University, Funabashi, 274-8510, Japan
}
\vspace*{0.5cm}
\\
$^{a}$mhatsuda@post.kek.jp
~~;~~
 $^{b}$kamimura@ph.sci.toho-u.ac.jp
\end{center}

\vspace*{1cm}

\centerline{\bf Abstract}
\vspace*{7mm}
We present ``M5 algebra" to derive 
Courant brackets of the generalized geometry of 
 $T\oplus \Lambda^2T^\ast \oplus \Lambda^5T^\ast$:
 The Courant bracket generates the generalized diffeomorphism including gauge transformations
of three and six form gauge fields.
The  Dirac bracket between selfdual gauge fields on a M5-brane 
gives a $C^{[3]}$-twisted contribution to the Courant brackets.
For M-theory compactified on
a  five dimensional torus the U-duality symmetry is SO(5,5) 
and the M5 algebra basis is in the 
16-dimensional spinor representation.
The M5 worldvolume diffeomorphism constraints can be written as
bilinear forms of  the basis and 
transform as a  SO(5,5) vector.
We also present an  extended space spanned by the 16-dimensional coordinates
with section conditions determined from
 the M5 worldvolume diffeomorphism constraints.
\vspace*{0.5cm}

 \vfill 

\thispagestyle{empty}
\setcounter{page}{0}
\newpage 
\section{Introduction}

U-duality symmetry is a powerful guiding principle to define M-theory.
Generalized manifold which extends diffeomorphism in T-duality covariant way
is introduced in \cite{Hitchin:2004ut} 
and it is applied  to  generalized geometry for M-theory \cite{Hull:2007zu,{Pacheco}}.
A manifest T-duality is realized  in \cite{Siegel:1993xq}
in a double field theory by duplicated coordinates. 
There are 
several approaches on the dualities;
in string mechanics \cite{Siegel:1993xq,Tsey},
in Gaillard-Zumino approach \cite{Duff:1989tf},
in string field theory \cite{Hull:2009mi}
and
in double field theory \cite{{Hohm:2010pp}}.  
D-branes and RR fields are key ingredients 
to promote 
T-duality to U-duality, 
\cite{{Hassan:1999bv},{Fukuma:1999jt},{Koerber:2005qi},{Albertsson:2008gq},{Hohm:2011zr},{Jeon:2012kd},{Hatsuda:2012uk},{Asakawa:2012px},{Kahle:2013kg}}.

Recent progress on U-duality \cite{Bonelli:2005ti,{Cederwall:2007je},{Coimbra:2011nw},{Thompson:2011uw},Aldazabal:2013mya,
{Berman:2010is},{Berman:2011pe},{Berman:2011cg},{Berman:2011jh},
{Hatsuda:2012vm},{Park:2013gaj}} has been  exploring new description of the M-theory and its geometry. 
Berman, Perry and collaborators reformulate the low energy effective theory of M-theory 
using a coset construction with manifest U-duality symmetry
 \cite{{Berman:2010is},{Berman:2011pe},{Berman:2011cg},{Berman:2011jh}}.
The U-duality symmetries of the 
M-theory with  $d$-dimensional   torus compactification 
depend on $d$.
It is  SL(5) for $d$=4 and its coset construction is presented in \cite{Berman:2010is}.
The M2-brane origin is clarified in a worldvolume approach \cite{Hatsuda:2012vm} 
and generalization of the Riemannian geometry is presented in \cite{Park:2013gaj}.
In this paper we clarify M5-brane origin of the SO(5,5) duality for $d$=5 compactification 
in which the supergravity action is reformulated by 
using coset construction of 
SO(5,5)/[SO(5)$\times$SO(5)]  \cite{Berman:2011pe}.

\par 
\medskip
 
In generalization of Riemannian geometry
the generalized diffeomorphism transformations are generated
by Courant bracket  in a duality covariant way.  
There are several approaches to construct such bracket;
integrability conditions of the  Dirac manifold structure  \cite{Courant}, 
exceptional group extension \cite{{Pacheco},{Aldazabal:2013mya}},
generalized 
derivatives including gauge fields \cite{{Bara},Berman:2010is,Berman:2011cg}
and a brane algebra 
\cite{Siegel:1993xq}.
The authors have used 
the brane algebra approach for Dp-branes \cite{Hatsuda:2012uk} and M2-brane \cite{Hatsuda:2012vm}. In this paper we extend it to M5-brane.
In the brane algebra approach 
an one parameter family of Courant brackets with parameter $K$ are derived.   
 Courant  bracket ($K=0$) is antisymmetric in two vectors, while 
Dorfman bracket ($K=1$) is not antisymmetric 
but gives gauge transformation rules directly.
Generators of a brane algebra are 
momentum and brane currents. For example
the ones for the string algebra are  $Z_M=(p_m,~\partial_\sigma x^m)$.
The brane algebra basis $Z_M$ is a  representation of the duality symmetry.
When the Hamiltonian constraint and worldvolume diffeomorphism constraints 
are written in terms of $Z_M$,
they should be also representation of the duality symmetry consistently. 
The generalized metric in the Hamiltonian is also a 
G/H coset representation in terms of  supergravity fields 
in which G is duality symmetry and H is its subgroup.

 \par 
\medskip 
To construct a theory with manifest duality symmetry 
a doubled space or an extended space is  introduced, where its coordinates are 
in the representation of duality symmetry. 
Section conditions are required to obtain the physical space. 
 It is natural to identify the section conditions with worldvolume diffeomorphism 
constraints of the probe brane,
since the brane algebra basis becomes 
the conjugate momentum of the extended space 
coordinate preserving brane constraints. 
In the string case the $\sigma$ diffeomorphism constraint is 
${\cal H}_\sigma=p_m\partial_\sigma x^m=
\frac{1}{2}Z_M\eta^{MN}Z_N=0$ 
with  O($d,d$) invariant metric $\eta^{MN}$.
The algebra basis   $Z_M$  is interpreted as  $\partial_M={\partial}/({\partial X^M})$,
conjugate to the doubled space coordinates $X^M$.
A physical space is obtained by the section condition 
$\triangle=\partial_M\eta^{MN}\partial_N=0$.

Section conditions for M-theory in $d$=4 are obtained  in 
\cite{Berman:2011cg} from the closure of the algebra of 
generalized  diffeomorphism including gauge transformation
\cite{Berman:2010is}. The extended coordinates are $y_{mn}=-y_{nm}$
in addition to the usual coordinates $x^m$ with $m=1,\cdots,4$.
They form SL(5) covariant coordinates $X^{\hat{m}\hat{n}}=
\left(
X^{m5}=x^m,~X^{mn}=\frac{1}{2}\epsilon^{mnl_1l_2}y_{l_1l_2}\right)$
 with $\hat{m}=(m,5)$.
The section conditions are
\bea
\epsilon^{\hat{m}\hat{m}_1\cdots\hat{m}_4 }\displaystyle\frac{\partial}{\partial X^{\hat{m}_1\hat{m}_2}}
\frac{\partial}{\partial X^{\hat{m}_3\hat{m}_4}}=0 \Leftrightarrow 
\left\{
\begin{array}{l}
\epsilon^{{m}_1\cdots{m}_4 }\displaystyle\frac{\partial}{\partial y_{{m}_1{m}_2}}
\frac{\partial}{\partial y_{{m}_3{m}_4}}=0\\
\displaystyle\frac{\partial }{\partial x^n}\frac{\partial}{\partial y_{mn}}=0
\end{array}\right.\label{sectionSL5}.
\eea
This is nothing but the diffeomorphism constraints 
for a M2-brane 
\cite{Hatsuda:2012vm}.
For a M5-brane in $d$=5 there is a scalar coordinate $y$
in addition to $(x^m,~y_{mn})$ with $m=1,\cdots,5$
to make 16-dimensional SO(5,5) spinor representation.
The section condition will be  modified from \bref{sectionSL5} to ones involving $y$.
In this paper we derive the section conditions for M-theory in $d$=5
from the M5-brane constraints. 
The 
section conditions  are a part of the BPS condition of eleven dimensional M-theory
\cite{West:2012qm}.
BPS D-branes and M-branes satisfy the BPS conditions of type II theories and M-theory
respectively as well as constraints.
There is a correspondence between 
the BPS projection and the $\kappa$-symmetry projection, 
which is roughly square root of bilinear constraints.
This is  a reflection of a correspondence between 
the global supersymmetry algebra and the local supersymmetry algebra
for a supersymmetric brane system.

\par 
\medskip
The organization of this paper is the following.
In section 2 we reformulate  M5-brane action given by Pasti, Sorokin and Tonin \cite{Pasti:1996vs}
in such a way that constraints of a M5-brane in the supergravity background
are written in bilinear forms of some basis $Z_M$.
The selfdual gauge field on a single M5-brane
is treated by Dirac bracket preserving worldvolume covariance. 
The M5 algebra is calculated and 
a series of Courant brackets for  $T\oplus \Lambda^2T^\ast \oplus \Lambda^5T^\ast$
 is obtained.
It is shown that the obtained Courant bracket
 gives correct generalized diffeomorphism including 
gauge transformations for $C^{[3]}$ and $C^{[6]}$.
In  section 3 we focus on M-theory in $d$=5, where U-duality symmetry is SO(5,5).
SO(5,5) spinor states are constructed
and the  SO(5,5)  and SO(5)$\times$SO(5) transformation rules are  presented. 
It is shown the worldvolume diffeomorphism constraints form a SO(5,5) vector,
while the Hamiltonian constraint determines 
 SO(5,5) transformation rules of 
the supergravity fields, $G_{mn}$ and $C_{mnl}^{[3]}$. 
In section 4 an extended space spanned by $X^M$ is presented with section conditions 
 determined from the M5 diffeomorphism constraints.
SO(5,5) covariant C-bracket is also presented. 
\par\vskip 6mm

\section{M5 algebra and Courant bracket}

A probe brane in supergravity background determines an algebra,
whose representation basis $Z_M$ satisfies following conditions:
\begin{enumerate}
\item{Transformation algebra generated by $Z_M$ is closed.}  
  \item {The  Hamiltonian constraint for a probe brane is bilinear form in $Z_M$ as 
  ${\cal H}_\perp=\frac{1}{2}Z_M{\cal M}^{MN}Z_N\approx 0$ where ${\cal M}^{MN}$
  is the  generalized metric.}
  \item{The set of worldvolume diffeomorphism constraints ${\cal H}_i\approx0$ can be written in bilinears  in $Z_M$ as $ Z_M\tilde{\rho}^{MN}Z_N\approx 0$, where $\tilde{\rho}^{MN}$ 
  is  a constant matrix.}
  \item{Charges of $Z_M$, such as momentum charge and brane charges,
   are rotated covariantly under U-duality. } 
  \end{enumerate} 
Once the brane algebra is found, Courant bracket 
is determined as an algebra between vectors in the space spanned by $Z_M$.

In this section local structure of the geometry generated by the
M5 algebra is presented.
 We begin with the M5 action given by Pasti, Sorokin and Tonin 
\cite{Pasti:1996vs} and we perform canonical analysis in  the temporal gauge
\cite{Bergshoeff:1998vx}. 
The selfdual condition on rank two  gauge fields are mixture of first class constraints and second class constraints. 
We do not fix the first class constraint, Gauss law constraint,
to preserve worldvolume five dimensional  covariance 
 necessary to compute M5 algebra and Courant bracket.
The second class constraints are treated  by using the Dirac bracket 
keeping the  worldvolume five dimensional covariance.
Using with the Dirac bracket the M5 algebra is obtained and
the Courant brackets for M5 is obtained. It gives 
correct generalized diffeomorphism including gauge transformations
of $C^{[3]}$ and its magnetic dual field  $C^{[6]}$.  

\subsection{M5 constraints}

We begin with an action for a M5 brane proposed by 
Pasti, Sorokin and Tonin \cite{Pasti:1996vs}
\bea
I&=&\displaystyle\int d^6\sigma~{\cal L}~~,~~{\cal L}~=~
{\cal L}_{DBI}+{\cal L}_{SD}+{\cal L}_{WZ},
\\
{\cal L}_{DBI}&=&-T
\sqrt{-h_{\tilde{\cal F}}  }~,~
h_{\cal F}=\det(h_{\hat{i}\hat{j}}+\tilde{{\cal F}}_{\hat{i}\hat{j}})~,~
h_{\hat{i}\hat{j}}=\partial_{\hat{i}}x^m\partial_{\hat{j}}x^n
G_{mn}~,~
\tilde{{\cal F}}_{\hat{i}\hat{j}}=h_{\hat{i}\hat{k}}h_{\hat{j}\hat{k}'}\tilde{{\cal F}}^{\hat{k}\hat{k}'},\nn\\
{\cal L}_{SD}&=&\displaystyle\frac{T\sqrt{-h}}{4}
\tilde{\cal F}^{\hat{i}\hat{j}}
{\cal F}_{\hat{i}\hat{j}\hat{k}}
n^{\hat{k}}~,~h=\det h_{\hat{i}\hat{j}}~~~~,~~~\hat{i}=0,1,\cdots,5,
\nn\\
{\cal L}_{WZ}&=&T
\epsilon^{\hat{i}_1\cdots \hat{i}_6}
\left(\frac{1}{6!}C^{[6]}_{\hat{i}_1\cdots \hat{i}_6}
+\frac{1}{2\cdot 3!^2}{\cal F}_{\hat{i}_1\hat{i}_2\hat{i}_3}
C^{[3]}_{\hat{i}_4\hat{i}_5 \hat{i}_6}
\right)~,\nn\\
{ F}_{\hat{i}\hat{j}\hat{k}}&=&\partial_{\hat{i}}A_{\hat{j}\hat{k}}
+\partial_{\hat{j}}A_{\hat{k}\hat{i}}
+\partial_{\hat{k}}A_{\hat{i}\hat{j}}~,~
{\cal F}_{\hat{i}\hat{j}\hat{k}}~=~{F}_{\hat{i}\hat{j}\hat{k}}
-\partial_{\hat{i}}x^m \partial_{\hat{j}}x^n \partial_{\hat{k}}x^l
C^{[3]}_{mnl}~, \nn
\eea
where  the rank two abelian gauge field $A_{\hat{i}\hat{j}}(\sigma)$ 
has DBI-type coupling 
with an 
auxiliary scalar variable $a(\sigma)$ as
\bea\tilde{{\cal F}}^{\hat{i}\hat{j}}&=&
\displaystyle\frac{1}{3!\sqrt{-h}}
\epsilon^{\hat{i}\hat{j}\hat{k}\hat{k}_1\hat{k}_2\hat{k}_3}
n_{\hat{k}}
{\cal F}_{\hat{k}_1\hat{k}_2\hat{k}_3}
~,~
n_{\hat{k}}~=~\displaystyle\frac{\partial_{\hat{k}}a}{
\sqrt{-h^{\hat{i}\hat{j}}\partial_{\hat{i}}a 
\partial_{\hat{j}}a}}~~~.
\nn
\eea

In the temporal gauge $a=\tau$ the vector $n_{\hat{i}}$ becomes
$n_{\hat{i}}=\delta_{\hat{i}}^0 (-h^{00})^{-1/2}$ and
$\tilde{\cal F}^{\hat{i}\hat{j}}$ becomes $\tilde{{\cal F}}^{0i}=0
$, $\tilde{{\cal F}}^{ij}=
\epsilon^{ijk_1k_2k_3}{\cal F}_{k_1k_2k_3}/({3! \sqrt{{\bf h}}})
$
with~${\bf h}=\det h_{ij}$~
and $\epsilon^{0 i_1\cdots i_5}\equiv \epsilon^{i_1\cdots i_5}$,
 ~$i=1,\cdots,5$
\cite{Bergshoeff:1998vx}.
Canonical momenta are defined as
\bea
p_m&=&\frac{\partial {\cal L}}{\partial (\partial_0 x^m)}\nn
~=~
\tilde{p}_m
-TG_{mn}t^n\label{momenta} \\
&&~~+T\left(
C_{m}^{[6]}
-\frac{\epsilon^{i_1\cdots i_5}
}{4!}(2{F}_{i_1i_2i_3}
-C^{[3]}_{i_1i_2i_3}
)\partial_{i_4}x^{m_1}\partial_{i_5}x^{m_2}
C^{[3]}_{m_1m_2m}
\right)\nn\\
E^{ij}&=&\frac{\partial {\cal L}}{\partial (\partial_0 A_{ij})}~
=~ \frac{T}{4}\epsilon^{iji_1i_2 i_3}\partial_{i_1}A_{i_2i_3}\nn\\
E^{0i}&=&\frac{\partial {\cal L}}{\partial (\partial_0 A_{0i})}~
=~0~,\nn
\eea
where ${\bf h}^{ij}$ is inverse of $h_{ij}$ and
\bea
\tilde{p}_m&=&\frac{\partial {\cal L}_{DBI}}
{\partial (\partial_0 x^m)}\label{ptilde}\\
t^n&=&
\partial_i x^n {\bf h}^{ij}
\frac{\epsilon^{i_1\cdots i_5}
}{4!}{\cal F}_{i_1i_2i_3}{\cal F}_{i_4i_5j}
~=~\partial_{i_5}x^n\frac{\epsilon^{i_1\cdots i_5}
}{8}\tilde{{\cal F}}_{i_1i_2}\tilde{{\cal F}}_{i_3i_4}
~=~\partial_ix^n\frac{1}{2T} {\bf h}^{ij}
{\cal F}_{ji_1i_2}\tilde{E}^{i_1i_2}
\nn\\
\tilde{E}^{ij}&=&\frac{2T}{4!}\epsilon^{iji_1i_2i_3}{\cal F}_{i_1i_2i_3}\nn\\
C^{[3]}_{i_1i_2i_3}
&=&\partial_{i_1}x^{n_1}\partial_{i_2}x^{n_2}\partial_{i_3}x^{n_3}
C^{[3]}_{n_1n_2n_3}~\nn\\
C_n^{[6]}&=&\frac{1}{5!}\epsilon^{i_1\cdots i_5}
\partial_{i_1}x^{n_1}\partial_{i_2}x^{n_2}
\partial_{i_3}x^{n_3}\partial_{i_4}x^{n_4}\partial_{i_5}x^{n_5}
C^{[6]}_{nn_1n_2n_3n_4n_5}~~~.
\eea

Hamiltonian constraints are given as
\bea
\left\{
{\renewcommand{\arraystretch}{1.4}
\begin{array}{ccl}
{\cal H}_\perp&=&\frac{1}{2T}
\left(
\tilde{p}_m G^{mn}\tilde{p}_n
+T^2\det (h+\tilde{\cal F})_{ij}
\right)~=~0\\
{\cal H}_i&=&\tilde{p}_m\partial_i x^m~=~0\\
\Phi^{ij}&=&\displaystyle\frac{1}{\sqrt{T}}E^{ij}-\frac{\sqrt{T}}{4}
\epsilon^{iji_1i_2 i_3}\partial_{i_1}A_{i_2i_3}=0
\end{array}}\right.\label{Hamcon}~~~.
\eea
Using an identity which holds for a 5-dimensional symmetric matrix  $S_{ij}=s_i{}^{\bar{a}}s_j{}^{\bar{a}}$ 
 and an antisymmetric matrix $T_{ij}=-T_{ji}$,  
with $i, \bar{a}=1,\cdots,5$, 
\bea
&\det(S_{ij}+T_{ij})=\det S_{ij}+\displaystyle\frac{1}{24}{\cal T}_1^{\bar{a}\bar{b}\bar{c}}{\cal T}_1^{\bar{a}\bar{b}\bar{c}}
+\displaystyle\frac{1}{64}{\cal T}_2^{\bar{a}}{\cal T}^{\bar{a}}_2&\label{2464},\\
&{\cal T}^{\bar{a}\bar{b}\bar{c}}_1=\epsilon^{i_1\cdots i_5}s_{i_1}{}^{\bar{a}}s_{i_2}{}^{\bar{b}}
s_{i_3}{}^{\bar{c}}T_{i_4i_5}~~,~~
{\cal T}^{\bar{a}}_2=\epsilon^{i_1\cdots i_5}s_{i_1}{}^{\bar{a}}T_{i_2i_3}T_{i_4i_5}&
\nn~~~
\eea
the determinant term in  ${\cal H}_\perp$ is rewritten as 
\bea
&&\det (h_{ij}+\tilde{{\cal F}}_{{i}{j}})\nn\\
&&~~=~\frac{1}{5!}\left(
\epsilon^{i_1\cdots i_5}\partial_{i_1}x^{m_1}\cdots
\partial_{i_5}x^{m_5}
\right)G_{m_1n_1}\cdots G_{m_5n_5}
\left(
\epsilon^{j_1\cdots j_5}\partial_{j_1}x^{m_1}\cdots
\partial_{j_5}x^{m_5}
\right)
\nn\\
&&~~~~~~+\frac{1}{24}
\left(\epsilon^{i_1\cdots i_5}\partial_{i_1}x^{m_1}
\partial_{i_2}x^{m_2}\partial_{i_3}x^{m_3}
\tilde{\cal F}_{i_4i_5}\right)
G_{m_1n_1}G_{m_2n_2}G_{m_3n_3}
\left(\epsilon^{j_1\cdots j_5}\partial_{j_1}x^{n_1}
\partial_{j_2}x^{n_2}\partial_{j_3}x^{n_3}
\tilde{\cal F}_{j_4j_5}\right)\nn
\\
&&~~~~~~+\frac{1}{64}
\left(\epsilon^{i_1\cdots i_5}\partial_{i_1}x^{m_1}
\tilde{\cal F}_{i_2i_3}
\tilde{\cal F}_{i_4i_5}\right)
G_{m_1n_1}
\left(\epsilon^{j_1\cdots j_5}\partial_{j_1}x^{n_1}
\tilde{\cal F}_{j_2j_3}
\tilde{\cal F}_{j_4j_5}\right)\nn\\
&&~~=~{5!}
(\frac{1}{5!}\epsilon^{i_1\cdots i_5}
\partial_{i_1}x^{m_1}
\cdots\partial_{i_5}x^{m_5})
 G_{m_1n_1}\cdots G_{m_5n_5} 
(\frac{1}{5!}\epsilon^{j_1\cdots j_5}
\partial_{j_1}x^{n_1}
\cdots\partial_{j_5}x^{n_5})
\nn\\
&&~~~~~~+\frac{2}{T^2}
(\tilde{E}^{i_1i_2}\partial_{i_1}x^{m_1}\partial_{i_2}x^{m_2})
G_{m_1n_1}G_{m_2n_2}
(\tilde{E}^{j_1j_2}\partial_{j_1}x^{n_1}\partial_{j_2}x^{n_2})
+t^mG_{mn}t^n~~~\label{ttdet}.
\eea
The diffeomorphism constraint  in \bref{Hamcon} leads to 
\bea
{\cal H}_i{\bf h}^{ij}{\cal F}_{ji_1i_2}{\cal F}_{i_3i_4i_5}\epsilon^{i_1\cdots i_5}/4!~=~
\tilde{\tilde{p}}_mt^m+T t^mG_{mn}t^n~=~0~~~,\label{diffeot}
\eea
where $\tilde{\tilde{p}}_m =\tilde{p}_m - TG_{mn}t^n$. 
The first term in the Hamiltonian in \bref{Hamcon} is rewritten by \bref{diffeot}
as 
\bea
\tilde{{p}}_mG^{mn}\tilde{{p}}_n
~=~\tilde{\tilde{p}}_mG^{mn}\tilde{\tilde{p}}_n
+2T\tilde{\tilde{p}}_mt^m+T^2t^mG_{mn}t^n~
=~\tilde{\tilde{p}}_mG^{mn}\tilde{\tilde{p}}_n
-T^2t^mG_{mn}t^n~\label{pptt}~~,
\eea 
where the $t^m$ dependent term is cancelled out 
with the second term, the contribution in the 5-dimensional 
spatial determinant  in \bref{ttdet}.
The basis for a M5 brane system is introduced as
\bea
&&Z_M~=~\left(
{\renewcommand{\arraystretch}{1.4}
\begin{array}{c}
Z_m\\
{Z}^{[2]}{}^{m_1m_2}\\
{Z}^{[5]}{}^{m_1\cdots m_5}
\end{array}}
\right)~=~
\left(
{\renewcommand{\arraystretch}{1.4}
\begin{array}{c}
p_m\\
2{E}^{i_1i_2}\partial_{i_1}x^{m_1}\partial_{i_2}x^{m_2}
\\
T\epsilon^{i_1\cdots i_5}
\partial_{i_1}x^{m_1}
\partial_{i_2}x^{m_2}\partial_{i_3}x^{m_3}\partial_{i_4}x^{m_4}
\partial_{i_5}x^{m_5}
\end{array}}
\right)
\label{ZMs}
\eea
As a result Hamiltonian constraints become
\bea
 \left\{\begin{array}{ccl}
{\cal H}_\perp&=&\displaystyle\frac{1}{2T}
\left(\tilde{\tilde{p}}_m G^{mn}\tilde{\tilde{p}}_n
+\frac{1}{2^2}\tilde{Z}^{[2]}{}^{m_1m_2}
G_{[m_1|n_1}G_{|m_2]n_2}\tilde{Z}^{[2]}{}^{n_1n_2}\right.\\
&&~~~~~\left.
+\displaystyle\frac{1}{5!^2}
{Z}^{[5]}{}^{m_1\cdots m_5}G_{[m_1|n_1}\cdots 
G_{|m_5]n_5}{Z}^{[5]}{}^{n_1\cdots n_5}\right)~=~0\\
{\cal H}_i&=&\tilde{\tilde{p}}_m\partial_i x^m
+\displaystyle\frac{1}{2}\tilde{E}^{jk}{\cal F}_{ijk}~=~p_m\partial_i x^m
+\displaystyle\frac{1}{2}{E}^{jk}{F}_{ijk}
\\
&=&p_m\partial_i x^m
+\displaystyle\frac{1}{2T}\epsilon_{ii_1\cdots i_4}
E^{i_1i_2}E^{i_3i_4}~=~0,
\end{array}\right.
\label{Hamconv2}
\eea
where
\bea
\tilde{\tilde{p}}_m&=&
p_m+\frac{1}{2}C^{[3]}_{mn_1n_2}Z^{[2]n_1n_2}
-\displaystyle\frac{1}{5!}(C^{[6]}_{mn_1\cdots n_5}
+5C^{[3]}_{mn_1n_2}C^{[3]}_{n_3n_4n_5})Z^{[5]n_1\cdots n_5}
\nn\\
\tilde{Z}^{[2]}{}^{n_1n_2}&=&
2\tilde{E}^{i_1i_2}\partial_{i_1}x^{n_1}\partial_{i_2}x^{n_2}
~=~{Z}^{[2]}{}^{n_1n_2}
-\frac{1}{3!}C^{[3]}_{m_1m_2m_3}Z^{[5]n_1n_2m_1m_2m_3}
\nn~~~.
\eea
The constraint ${\cal H}_\perp$  can be written in terms of bilinear form of
M5 brane basis, $Z_M$ in \bref{ZMs},
\bea
&&{\cal H}_\perp=\frac{1}{2T}
Z_M{\cal M}^{MN}Z_N\label{Hamperp}\\
&&~~{\cal M}^{MN}= ({\cal N}^T)^M{}_L{\cal M}_0{}^{LK}{\cal N}_K{}^N\nn\\
&&~~
{\cal M}_0{}^{ML}=\left(
{\renewcommand{\arraystretch}{1.4}
\begin{array}{ccc}
G^{ml}&0&0\\
0&G_{[m_1|l_1}G_{|m_2]l_2}&0\\
0&0&G_{[m_1|l_1}\cdots G_{|m_5]l_5}
\end{array}}
\right)\label{M0M0}\\
&&~~{\cal N}_L{}^N=
\left(
{\renewcommand{\arraystretch}{1.4}
\begin{array}{ccc}
\delta_l^{n}&C^{[3]}_{ln_1n_2}&
-C^{[6]}_{ln_1\cdots n_5}-
\displaystyle\frac{1}{4!}C^{[3]}_{l[n_1n_2}C^{[3]}_{n_3n_4n_5]}\\
0&\delta^{l_1}_{[n_1}\delta^{l_2}_{n_2]}&
-\displaystyle\frac{1}{3!}C^{[3]}_{[n_1n_2n_3}\delta_{n_4}^{l_1}
\delta_{n_5]}^{l_2}
\\
0&0&\delta^{l_1}_{[n_1}\cdots \delta^{l_5}_{n_5]}
\end{array}}
\right)\nn~~~.
\eea
Indices are contracted as $U^MV_M=U^mV_m+\frac{1}{2}U^{m_1m_2}V_{m_1m_2}
+\frac{1}{5!}U^{m_1\cdots m_5}V_{m_1\cdots m_5}$.

Worldvolume spatial diffeomorphism constraints ${\cal H}_i=0$ are
also written in bilinears of $Z_M$ by contracting with
$\epsilon^{ii_1\cdots i_4}\partial_{i_1}x^{m_1}\cdots
\partial_{i_4}x^{m_4}$ and $E^{ij}\partial_{j}x^{m}$,
\bea
{\cal H}_i=0
&\Rightarrow&
Z_M\tilde{\rho}^{MN}Z_N=
(Z_M \tilde{\rho}_{[M2]}^{MN} Z_N)^n~a_n+
(Z_M\tilde{\rho}_{[M5]}^{MN}Z_N)^{n_1\cdots n_4}~b_{n_1\cdots n_4}=
0~~,\label{diffeoZZ}\\
&&
\tilde{\rho}^{MN}=
\left(
{\renewcommand{\arraystretch}{1.4}
\begin{array}{ccc}
0&a_{[n_1}\delta^m_{n_2]}&b_{[n_1\cdots n_4}\delta^m_{n_5]}\\
a_{[m_1}\delta^n_{m_2]}&b_{[m_1m_2n_1n_2]}&0\\
b_{[m_1\cdots m_4}\delta^n_{m_5]}&0&0
\end{array}}
\right),
\nn
\eea
where $a_m$ and $b_{m_1\cdots m_4}$ are arbitrary constants.
Not only the M5-brane diffeomorphism constraints
but also  M2-brane diffeomorphism constraints obtained in 
\cite{Hatsuda:2012vm} appear. 

A single M5-brane system including the selfdual gauge field
  reduces into a single D4-brane system including a Dirac-Born-Infeld gauge field
by the  double dimensional reduction  
\cite{Perry:1996mk,Ho:2008ve}.
The set of basis $Z_M$ for a M5 brane 
given in \bref{ZMs} 
corresponds to 
the one for a D4 brane in the IIA theory given in 
\cite{{Hatsuda:2012uk}}
as
\bea
{\renewcommand{\arraystretch}{1.4}
\begin{array}{ccccc}
&{\rm M5}&\to&{\rm D4}&\\ \hline
{\rm momentum}&Z_m&&p_m&{\rm momentum}\\
{\rm M2~mode}&Z^{[2]}{}^{mn}&&\frac{1}{2}E^{i}\partial_{i}x^{m}&
{\rm string~mode} \\
{\rm momentum~on~M5}&t^{m} &&\frac{1}{4}\epsilon^{j_1j_2k_1k_2}
{F}_{j_1j_2}{F}_{k_1k_2}&{\rm D0~mode~on~D4}\\
{\rm M2~mode~on~M5}&Z^{[2]}{}^{m_1m_2}&&\epsilon^{i_1i_2j_1j_2}
{F}_{i_1i_2}\partial_{j_1}x^{m_1}\partial_{j_2} x^{m_2}&{\rm D2~mode~on~D4}\\
{\rm M5~mode}&Z^{[5]}{}^{m_1\cdots m_5}&&\frac{1}{4!}\epsilon^{i_1\cdots i_4}
\partial_{i_1}x^{m_1}\cdots\partial_{i_4} x^{m_4}&{\rm D4~mode}
\end{array}}~~~\nn\\
\eea
The M2 mode  $Z^{[2]}$ on M5 reduces into both string mode and D2 mode on D4
 from  selfdual property.
The M5 algebra basis reduces into the one for D4 algebra
except the uplifted D0 mode on D4, whose origin $t^m$ appears in the momentum $\tilde{p}_m$
only at the first stage.

\subsection{Selfdual gauge field}

A  M5 brane includes
 M2 brane boundaries which are
selfdual rank two antisymmetric gauge field.
The gauge field $A_{ij}$ and its  canonical conjugate $E^{ij}$ satisfy
the Poisson bracket 
\bea
\{E^{ij}(\sigma),A_{i'j'}(\sigma')\}
&=&-i\delta^{[i}_{i'}\delta^{j]}_{j'}\delta(\sigma-\sigma') ~
\eea 
as well as the selfduality constraints in \bref{Hamcon}, $\Phi^{ij}=0$.
This  conditions is a mixture of first class and second class constraints \cite{Bengtsson:1996fm}.
Its longitudinal modes are first class 
Gauss law constraints,
 $\partial_i\Phi^{ij}=\partial_iE^{ij}=0$
and the transverse modes are  second class 
\bea
\Phi_\perp{}^{ij}&=&{\cal P}_\perp{}^i_{i'}{\cal P}_\perp{}^j_{j'}\Phi^{i'j'}~=~0~~,~~
{\cal P}_\perp{}_i^{j}=\delta_i^j-\frac{\partial_i \partial^j}{\bigtriangleup }
\eea
satisfying 
\bea
\left\{
\Phi_\perp{}^{j_1j_2}(\sigma),\Phi_\perp{}^{j_3j_4}(\sigma')
\right\}=
-i\epsilon^{j_1j_2j_3j_4i}\partial_i\delta(\sigma-\sigma')
&\equiv&
\Xi^{j_1j_2j_3j_4}(\sigma,\sigma')~.
\label{invers}\eea
The  inverse of \bref{invers} exists only in the transverse directions
\bea
&(\Xi^{-1})_{i_1i_2i_3i_4}(\sigma,\sigma')
=i\epsilon_{i_1i_2i_3i_4i}
\displaystyle\frac{1}{\bigtriangleup }\partial^i
\delta(\sigma-\sigma'),&\nn\\
&\displaystyle\frac{1}{2}\int d\sigma'(\Xi^{-1})_{ij~i'j'}(\sigma,\sigma')(\Xi)^{i'j'~kl}(\sigma',\sigma'')
={\cal P}_\perp{}_i^{[k}{\cal P}_\perp{}_j^{l]}\delta(\sigma-\sigma'')&\nn~~.
\eea
The Dirac bracket, verifying 
$\left\{\Phi_\perp{}^{ij}(\sigma),{\cal O}(\sigma')\right\}_D=
0$ for any ${\cal O}$ , is defined as
\bea
&&\left\{{\cal O}_1(\sigma),{\cal O}_2(\sigma')\right\}_D~=~
\left\{{\cal O}_1(\sigma),{\cal O}_2(\sigma')\right\}\nn\\
&&~~~~~-\displaystyle\int d\sigma^{''} 
\left\{{\cal O}_1(\sigma),\Phi^{ij}(\sigma^{''})\right\}
\frac{i}{4}\epsilon_{iji'j'k}\frac{1}{\bigtriangleup'' }\frac{\partial}{\partial
\sigma''_k}
\left\{\Phi^{i'j'}(\sigma^{''}),{\cal O}_2(\sigma')\right\}\label{Dirac}
~~~.\nn
\eea
It leads to nontrivial bracket between two  $E^{ij}$'s as
\bea
\left\{
E^{i_1i_2}(\sigma),E^{i_3i_4}(\sigma')
\right\}_D&=&\frac{iT}{4}
\epsilon^{i_1i_2i_3i_4k}\partial_{k}\delta(\sigma-\sigma')
~~~.\label{EEDirac}
\eea
\par

\subsection{M5 algebra and Courant brackets}

Now let us compute the M5 algebra.
The Dirac bracket between $Z_M$'s  in \bref{ZMs}
 is given by
\bea
&\left\{Z_M(\sigma),Z_N(\sigma')\right\}_D~=~
iT\rho_{MN}^i\partial_i\delta(\sigma-\sigma'),&\nn\\
\nn\\
&\rho^i_{MN}~=~\left(
{\renewcommand{\arraystretch}{1.4}
\begin{array}{ccc}
0&\frac{1}{T}E^{ji}\partial_jx^{[n_1}\delta^{n_2]}_m
&r_{~m}^{i;n_1\cdots n_5}
\\
\frac{1}{T}E^{ji}\partial_jx^{[m_1}\delta^{m_2]}_n
& r^{i;m_1m_2n_1n_2l}_{~l}&0
\\r_{~n}^{i;m_1\cdots m_5}&0&0
\end{array}}
\right)&,\nn\\
&r^{i;n_1\cdots n_5}_{~m}=
\frac{1}{4!}\epsilon^{ii_1\cdots i_4}
\partial_{i_1}x^{[n_1}
\cdots \partial_{i_4}x^{n_4}\delta^{n_5]}_m~,
&\label{rhoM5}
\eea
where \bref{EEDirac} is used.
The matrix $\rho_{MN}$ is symmetric and satisfies the following relations
\bea
\partial_i\rho_{MN}^i=0~~,~~
\rho_{MN}^i\partial_ix^l\partial_l=f_{MN}^{L;m}Z_L\partial_m~~~
\label{rhof}
\eea
by the Gauss law constraint.
We introduce local vector $\Lambda^M(\sigma)$  in the M5 basis  \bref{ZMs} as 
\bea
&\hat{\Lambda}(\sigma)~=~\Lambda^MZ_M~=~\lambda+\lambda^{[2]}+\lambda^{[5]}
\in T \oplus \Lambda^2 T^* \oplus \Lambda^5 T^*&\nn\\
&\left\{
{\renewcommand{\arraystretch}{1.4}
\begin{array}{rcl}
\lambda&=&\Lambda^mZ_m \\
\lambda^{[2]}&=&\frac{1}{2}\Lambda_{m_1m_2}Z^{[2]m_1m_2}
\\
\lambda^{[5]}&=&\frac{1}{5!}\Lambda_{m_a\cdots m_5} Z^{[5]m_a\cdots m_5}
\end{array}}\right. ~~.&
\eea
The M5 algebra is closed 
up to the brane-like anomalous terms as
\bea
\left\{\hat{\Lambda}_1(\sigma)
,\hat{\Lambda}_2(\sigma')\right\}_D&=&
-iT\hat{\Lambda}_{12}\delta(\sigma-\sigma')\nn\\&&~~
+iT\left(
\frac{1-K}{2}\Psi^i_{(12)}(\sigma)+
\frac{1+K}{2}\Psi^i_{(12)}(\sigma')
\right)\partial_i\delta(\sigma-\sigma')\nn\\
\label{L12}\\
\hat{\Lambda}_{12}
&=&\Lambda_{[1}^n\partial_n\Lambda_{2]}^MZ_M
-\frac{1}{2}\Lambda_{[1}^M\rho_{MN}^i\partial_i\Lambda_{2]}^N
+\frac{K}{2}\partial_i\Psi^i_{(12)}\nn\\
&=&\left(
\Lambda_{[1}^n\partial_n\Lambda_{2]}^M
-\frac{1}{2}f_{LN}^{M;n}
\Lambda_{[1}^L \partial_n\Lambda_{2]}^N
+\frac{K}{2}
f_{LN}^{M;n}
\Lambda_{(1}^L \partial_n\Lambda_{2)}^N
\right)Z_M
\nn\\
\Psi^i_{(12)}&=& \Lambda_1^M\rho_{MN}^i\Lambda_{2}^N=
\frac{1}{2} \Lambda_{(1}^M\rho_{MN}^i\Lambda_{2)}^N~.
 \nn
\eea
$K$ is an arbitrary constant reflecting 
an ambiguity of $\partial_i\delta(\sigma-\sigma')$ term as shown in \cite{Hatsuda:2012uk}.
The vector  $\hat{\Lambda}_{12}$ in \bref{L12} is
recognized as one parameter family 
of Courant bracket for $ T \oplus \Lambda^2 T^* \oplus \Lambda^5 T^*$ 
derived from the M5 algebra
\bea
\left[\hat{\Lambda}_1,\hat{\Lambda}_2
\right]_{M5,K}=
\left(
\Lambda_{[1}^n\partial_n\Lambda_{2]}^M
-\frac{1}{2}f_{LN}^{M;n}
\Lambda_{[1}^L \partial_n\Lambda_{2]}^N
+\frac{K}{2}
f_{LN}^{M;n}
\Lambda_{(1}^L \partial_n\Lambda_{2)}^N
\right)Z_M \label{CM5K}~~~.
\eea
For $K=0$ the bracket is the Courant bracket which is antisymmetric in 
$\Lambda_1 \leftrightarrow \Lambda_2$:
\bea
&&\left[\hat{\Lambda}_1,\hat{\Lambda}_2
\right]_{M5,K=0}=
\left[\lambda_1,\lambda_2\right]
+{\cal L}_{\lambda_{[1}}\lambda^{[2]}_{2]}
+{\cal L}_{\lambda_{[1}}\lambda^{[5]}_{2]}
-\frac{1}{2}d(\iota_{\lambda_{[1}}\lambda^{[2]}_{2]})
-\frac{1}{2}d(\iota_{\lambda_{[1}}\lambda^{[5]}_{2]})\nn\\
&&~~~~~~~~~~~~~~~~~~-\frac{1}{2}\lambda^{[2]}_{[1}\wedge d\lambda_{2]}^{[2]}
\\
&&~~~~
\left\{
\begin{array}{ccl}
[\lambda_1,\lambda_2]&=&\Lambda^n_{[1}\partial_n\Lambda^m_{2]}Z_m\\
{\cal L}_{\lambda_1}\lambda^{[p]}_{2}&=&
\frac{1}{p!}\left(
\Lambda_1{}^l\partial_l \Lambda_2{}^{[p]}_{m_1\cdots m_p}
+\partial_{[m_1}\Lambda_1{}^{l}\Lambda_2{}^{[p]}_{|l|m_2\cdots m_p]}\right)
Z^{[p]m_1\cdots m_p}\label{CCourant}\\
d(\iota_{\lambda_{1}}\lambda^{[p]}_{2})
&=&\frac{1}{p!}\partial_{[m_1|}
\left(
\Lambda_{1}{}^l\Lambda_{2;}{}_{l|m_2\cdots m_p]}
\right)Z^{[p]m_1\cdots m_p}~.
\end{array}\right.
\nn
\eea
This result is consistent with the one in \cite{{Pacheco}}.
A useful choice is  $K=1$ and   
the bracket is Dorfman bracket. It
is not antisymmetric in $\Lambda_1 \leftrightarrow \Lambda_2$,
but it gives totally antisymmetrized gauge transformations of 
antisymmetric gauge fields automatically. 
\bea
&&\left[\hat{\Lambda}_1,\hat{\Lambda}_2
\right]_{M5,K=1}=
\left[\lambda_1,\lambda_2\right]
+{\cal L}_{\lambda_{1}}\lambda^{[2]}_{2}
+{\cal L}_{\lambda_{1}}\lambda^{[5]}_{2}
-\iota_{\lambda_{2}}d\lambda_1^{[2]}
-\iota_{\lambda_{2}}d\lambda_1^{[5]}\nn\\
&&~~~~~~~~~~~~~~~~~~+
\lambda^{[2]}_{2}\wedge 
d\lambda_{1}^{[2]}\label{Dorf}
\eea
with
\bea
&&~~~~
\iota_{\lambda_{2}}d\lambda_1^{[p]}
=\frac{1}{p!^2}\Lambda_{2}{}^l
\partial_{[l|}
\Lambda_1{}_{|m_1\cdots m_p]}
Z^{[p]m_1\cdots m_p}
\nn~~~.
\eea

Using with the obtained Courant bracket for $K=1$, Dorfman bracket in \bref{Dorf}, 
the general coordinate transformation compatible with the M5 brane background 
 is given as 
\bea
&&
\delta_\xi \hat{E}^{M5}_a~=~\left[\hat{\xi},\hat{E}_a^{M5}\right]_{M5,K=1}~~\nn\\
&&(\hat{E}_a^{M5})^M=\left(
{\renewcommand{\arraystretch}{1.4}
\begin{array}{c}
e_a{}^m\\
e_a{}^mC_{mm_1 m_2}^{[3]}\\
e_a{}^mC_{mm_1 \cdots m_5}^{[6]}
\end{array}}
\right)~~,~~
(\hat{\xi})^M=\left(
{\renewcommand{\arraystretch}{1.4}
\begin{array}{c}
\xi{}^m\\
\xi_{m_1 m_2}^{[2]}\\
\xi_{m_1 \cdots m_5}^{[5]}
\end{array}}
\right)~~.
\eea
By contracting the local Lorentz SO(5) index $a$, $e_m{}^ae_a{}^n=\delta_m^n$
and $e_m{}^ae_n{}^b\delta_{ab}=G_{mn}$,
it gives the expected general coordinate transformations and gauge transformations 
as
\bea
\left\{
{\renewcommand{\arraystretch}{1.4}
\begin{array}{lcl}
\delta_\xi G_{mn}&=&\xi^l\partial_lG_{mn}
+\partial_{(m|}\xi^l G_{l|n)}
\\
\delta_\xi C_{m_1m_2m_3}^{[3]}&=&
\xi^l\partial_lC_{m_1m_2m_3}^{[3]}
+\frac{1}{2}\partial_{[m_1|}\xi^l C_{l|m_2m_3]}^{[3]}
-\frac{1}{2}\partial_{[m_1}\xi^{[2]}_{m_2m_3]}
\\
\delta_\xi C_{m_1\cdots m_6}^{[6]}&=&
\xi^l\partial_lC_{m_1\cdots m_6}^{[6]}
+\frac{1}{5!}\partial_{[m_1|}\xi^l C_{l|\cdots m_6]}^{[6]}
-\frac{1}{5!}\partial_{[m_1}\xi^{[5]}_{\cdots m_6]}
+\frac{1}{4!}C^{[3]}_{[m_1m_2m_3}\partial_{m_4}\xi_{m_5m_6]}^{[2]}~.
\end{array}}\nn
\right.\\
\eea
The last term in the  gauge transformation of $C^{[6]}$ 
comes from the Dirac bracket between selfdual gauge fields.

\section{SO(5,5) duality symmetry}

We focus on a M-theory in $d$=5, where the duality symmetry group is SO(5,5). 
The SO(5,5;Z) duality mixes quantized  momentum,  
M2 and M5-brane charges, while SO(5,5;R) duality 
symmetry rotates momentum and M2 and M5-brane currents in the
low energy effective theory. 
In this section we consider SO(5,5;R) transformation of the 
M5 algebra basis  \bref{ZMs}, which is
 16-dimensional spinor representation.
The supergravity fields, metric and three form gauge field,
are coset parameters of G/H with G=SO(5,5) and  H=SO(5)$\times$SO(5).

\par

\subsection{SO(5,5) spinor states}

In the 
case of $d$=5 the five form $Z^{[5]m_1\cdots m_5}$ 
becomes a pseudo scalar $Z^{[5]}$,
then the M5 basis becomes $5+10+1$=16-dimensional spinor representation 
of SO(5,5), 
\bea
Z_M=\left(
Z_m,~Z^{[2]m_1m_2},~Z^{[5]}
\right)~~~,~m=1,\cdots, 5~.
\eea
The 16-dimensional spinor representation 
of SO(5,5) duality group is constructed by  fermionic oscillators 
$\psi^m$ and $\psi^\dagger_m$  \cite{Fukuma:1999jt}
\bea
&\left\{\psi^\dagger_m,~\psi^n\right\}=\delta_{m}^n~~,~~
\left\{\psi^m,~\psi^n\right\}=
\left\{\psi^\dagger_m,~\psi^\dagger_n\right\}=0~~.&
\eea
 The SO(5,5) Clifford algebra is 
\bea
&&\Gamma_{\hat{a}}=
(\Gamma_a,~\Gamma_{\dot{a}})=
\left\{
\begin{array}{l}
\Gamma_a=\psi^me_{ma}+e_a{}^m\psi^\dagger_m\\
\Gamma_{\dot{a}}=\psi^me_{m}{}_a-e_{a}{}^{m}\psi^\dagger_m
\end{array}\right.~~
a,\dot{a}=1,\cdots,5
\\
&&\left\{\Gamma_{\hat{a}},\Gamma_{\hat{b}}
\right\}=\hat{\eta}_{\hat{a}\hat{b}}~~,~~
\hat{\eta}_{\hat{a}\hat{b}}={\rm diag}(\delta_{ab},-\delta_{\dot{a}\dot{b}})
\nn\\
&&\Gamma_{11}=
\displaystyle\prod_{\hat{a}}\Gamma_{\hat{a}}
=\prod_{m=1}^5 \left[\psi_m, \psi^\dagger_m\right]
~.\nn
\eea
The SO(5,5) chiral spinor states are 
constructed by acting odd
numbers of fermions  
on $ |+\rangle$ defined by  $\psi^{{m}} |+\rangle=0$ 
for a choice $\Gamma_{11}=-1$. 
They are 
\bea
|M\rangle&=&\left(
|_m\rangle~,~|^{m_1m_2}\rangle~,~|-\rangle~
\right)\nn \\&=&
\left\{
{\renewcommand{\arraystretch}{1.4}
\begin{array}{lcl}
|_m\rangle&=&
\psi^\dagger_{m}|+\rangle\\
|^{m_1m_2}\rangle&=&
\frac{1}{3!}\epsilon^{m_1m_2m_3m_4m_5}
\psi^\dagger_{m_3}\psi^\dagger_{m_4}\psi^\dagger_{m_5}
|+\rangle\\
|-\rangle&=&
\frac{1}{5!}\epsilon^{m_1m_2m_3m_4m_5}
\psi^\dagger_{m_1}\psi^\dagger_{m_2}
\psi^\dagger_{m_3}\psi^\dagger_{m_4}\psi^\dagger_{m_5}
|+\rangle
\end{array}}\right.~~~,\nn\\
\langle N|M\rangle&=&\hat{\delta}_{NM}~=~
\left(
\begin{array}{ccc}
\delta^m_n&0&0\\
0&\delta^{n_1}_{[m_1}\delta^{n_2}_{m_2]}&0\\
0&0&1
\end{array}
\right).\label{1616}
\eea

Infinitesimal SO(5,5) rotations are parameterized by   
 $\alpha_n{}^m,~\beta_{[nm]},~\gamma^{[nm]}$ as 
\bea
&&{\rm SO(5,5)}\ni g~~,~~g^T \eta g=\eta
~~,~~\eta=\left(
\begin{array}{cc}
0&{\bf 1}\\{\bf 1}&0
\end{array}
\right)
\nn\\
&&~~~~~\Rightarrow~
g_{\hat{m}}{}^{\hat{n}}=
\left(
{\renewcommand{\arraystretch}{1.4}
\begin{array}{cc}
(1+\alpha)_m{}^n&\beta{}_{mn}\\
\gamma^{mn}&(1+\alpha)_n{}^m
\end{array}}
\right)
~,~\beta_{mn}=-\beta_{nm}~,~
\gamma{}_{mn}=-\gamma{}_{nm}~.
 \label{frep}\eea
The chiral spinor  states $|M\rangle$ 
are transformed as
\bea
&&\delta |M\rangle ~=~\hat{S}|M\rangle~
=~S_M{}^N|N\rangle,\qquad ~\hat{S}=
\frac{1}{2}\alpha_n{}^m
[\psi^\dagger_m,\psi^n]+\frac{1}{2}\beta_{nm}
\psi^m\psi^n
+\frac{1}{2}\gamma^{nm}\psi^\dagger_m\psi^\dagger_n, \nn\\
&&\nn\\
&&S_M{}^{N}=
\left(
{\renewcommand{\arraystretch}{1.4}
\begin{array}{ccc}
-\frac{\hat{\alpha}}{2}\delta_n^m+\alpha_n{}^m&
-\tilde{\gamma}_{nm_1m_2}&0\\
\tilde{\beta}^{n_1n_2m}&
\frac{\hat{\alpha}}{2}\delta_{[m_1}^{n_1}
\delta_{m_2]}^{n_2}
-\alpha_{[m_1}{}^{[n_1}\delta_{m_2]}^{n_2]}&-\gamma^{n_1n_2}\\
0&\beta_{m_1m_2}&\frac{\hat{\alpha}}{2}
\end{array}}
\right),\label{SO55S}
\eea
where
$\hat{\alpha}=\alpha_m{}^m$,
~$\tilde{\beta}^{n_1n_2m}=\epsilon^{n_1n_2mk_1k_2}\beta_{k_1k_2}/2$. 

Under typical T-duality rotations a SO(5,5) vector $(v_m,\tilde{v}^m)$ is transformed using 
 $g$ in \bref{frep} as 
 $\delta v_m=\beta_{mn}\tilde{v}^n$ and  $\delta \tilde{v}^m=\gamma^{mn}{v}_n$.
 The M5 basis, $Z_M$, is transformed in the same way as the SO(5,5) spinor states 
$|M\rangle$ as in \bref{SO55S},
\bea
&
\delta p_m=-\frac{1}{4}\epsilon_{mn_1\cdots n_4}\gamma^{n_1n_2}Z^{[2]}{}^{n_3n_4}~,~
\delta Z^{[5]}=\frac{1}{2}\beta_{mn}Z^{[2]}{}^{mn}&\nn\\
&\delta Z^{[2]}{}^{mn}=\frac{1}{2}\epsilon^{mnl_1 l_2 l}\beta_{l_1l_2}p_{l}-
\gamma^{mn}Z^{[5]}~~~.&
\eea
SO(5,5) contains a subgroup  SO(5)$\times$SO(5) 
 generated by  
\bea
\frac{1}{4}\omega_+{}^{ab}\Gamma_{[a}\Gamma_{b]} ~+~
\frac{1}{4}\omega_-{}^{ab}
\Gamma_{[\dot{a}}\Gamma_{\dot{b}]}~~,~~
\left\{\begin{array}{l}
\omega_+{}^{ab}=\omega^{ab}+\bar{\omega}^{ab}\\
\omega_-{}^{\dot{a}\dot{b}}=\omega^{ab}-\bar{\omega}^{ab}
\end{array}\right.\label{rlw}
\eea
with infinitesimal parameters $\omega^{ab}$ and $\bar{\omega}^{ab}$,
\bea
&&
{\rm SO(5)} \times {\rm SO(5)}\ni h~~,~~
h^T \eta h=\eta ~~,~ ~
h^T \hat{\delta} h=\hat{\delta}~\nn\\
&&~~~~~\Rightarrow~
h_{\hat{a}}{}^{\hat{b}}=
\left(
{\renewcommand{\arraystretch}{1.4}
\begin{array}{cc}
(1+\omega)_a{}^b&\bar{\omega}{}_{ab}\\
\bar{\omega}{}^{ab}&(1+\omega)^a{}_b
\end{array}}
\right)\label{omega}
~,~\omega_{ab}=-\omega_{ba}~,~
\bar{\omega}{}_{ab}=-\bar{\omega}{}_{ba}~~.\nn
\eea
Under the SO(5)$\times$SO(5) transformation  spinor states 
are transformed as
\bea
&\delta_h|A\rangle
=S_h{}_A{}^{B}|B\rangle~~,~~
S_h{}_{A}{}^{B}=
\left(
{\renewcommand{\arraystretch}{1.4}
\begin{array}{ccc}
\omega_a{}^b&-\tilde{\bar{\omega}}_{ab_1b_2}&0\\
\tilde{\bar{\omega}}^{a_1a_2b}&
-\omega_{[b_1}{}^{[a_1}\delta_{b_2]}^{a_2]}&-\bar{\omega}^{a_1a_2}\\
0&\bar{\omega}_{b_1b_2}&0
\end{array}}
\right)~~~&\label{SO5SO5S}
\eea
with 
$|A\rangle $
$=\nu_A{}^M |M\rangle$ 
for a coset element $\nu\in \,$SO(5,5)/[SO(5)$\times$SO(5)].
This is analogous to right/left separation of string modes.
For the right mover SO(5)  transformations are given 
 with the parameter $\omega_+$ in \bref{rlw} as
 \bea
&\delta (V_a+\tilde{V}_{a})=
(\omega_+)_{ab}(V_b+\tilde{V}_{b}) &\nn\\
&V_a=p_aZ^{[5]}-\displaystyle\frac{1}{8}\epsilon_{aa_1\cdots a_4}Z^{[2]}{}^{a_1a_2}Z^{[2]}{}^{a_3a_4}
~,~\tilde{V}^a=p_b Z^{[2]}{}^{ba}~,\label{right}&
\eea where
$Z_A=\nu_A{}^MZ_M$.
The left mover is transformed similarly by replacing the $\pm$ signs.  
\par

\subsection{SO(5,5) transformation of M5 constraints}

For a probe M5-brane system  worldvolume diffeomorphism 
constraints must be covariant
under the SO(5,5) transformations.
  Diffeomorphism constraints ${\cal H}_i=0$ in \bref{Hamconv2}
 are  recasted into the ones for M2-brane and M5-brane as
in \bref{diffeoZZ}.
In $d$=5 
M5-brane diffeomorphism constraints 
$(Z_M \tilde{\rho}_{[M5]}^{MN} Z_N )_{m_1\cdots m_4}\approx 0$
become a five dimensional vector. Together with the one for M2
it forms a SO(5,5) fundamental  vector multiplet 
\bea
\left\{
{\renewcommand{\arraystretch}{1.4}
\begin{array}{lcl}
(Z\tilde{\rho}_{[M5]}Z)_m&=&2Z_mZ^{[5]}-
\displaystyle\frac{1}{4}\epsilon_{mm_1\cdots m_4}Z^{[2]m_1m_2}Z^{[2]m_3m_4}\approx 0\\
(Z\tilde{\rho}_{[M2]}Z)^m&=&2Z_nZ^{[2]mn}\approx 0
\end{array}}\right.~~~.
\eea
Under the SO(5,5) transformation given in \bref{SO55S}
\bea
&&\left\{
{\renewcommand{\arraystretch}{1.4}
\begin{array}{lcl}
\delta Z_m&=&\alpha_m{}^nZ_n-\frac{\hat{\alpha}}{2}Z_m-\frac{1}{2}
\tilde{\gamma}_{mn_1n_2}Z^{[2]n_1n_2}\\
\delta Z^{[2]m_1m_2}&=&\tilde{\beta}^{m_1m_2n}Z_n+
\alpha_n{}^{[m_1}Z^{m_2]n}+\frac{\hat{\alpha}}{2}Z^{[2]m_1m_2}
-{\gamma}^{m_1m_2}Z^{[5]}\\
\delta Z^{[5]}&=&\frac{1}{2}\beta_{n_1n_2}Z^{[2]n_1n_2}
+\frac{\hat{\alpha}}{2}Z^{[5]}
\end{array}}\right.
\nn~~,
\eea
they make a covariant SO(5,5)
vector,  which are consistent with the constraints:
\bea
&&\delta (Z\tilde{\rho}_{[M5]}Z){}_m~=~
\alpha_m{}^n (Z\tilde{\rho}_{[M5]}Z)_n
+\beta_{mn}(Z\tilde{\rho}_{[M2]}Z)^n
\nn\\
&&\delta(Z\tilde{\rho}_{[M2]}Z)^m~=~-\alpha_n{}^m(Z\tilde{\rho}_{[M2]}Z)^n
-\gamma^{mn}(Z\tilde{\rho}_{[M5]}Z){}_n. \nn
\eea
It is interesting to notice that the M5 diffeomorphism constraints
$(Z\tilde{\rho}_{[M5]}Z)_m=0$ 
is equal to $\tilde{p}_m=\tilde{\tilde{p}}_m+TG_{mn}t^n\approx 0$
 in \bref{ptilde}. 

Now let us examine the SO(5,5) transformations of the supergravity background
fields.
The Hamiltonian constraint in \bref{Hamperp} is 
written, for  5-dimensional part,   as 
\bea
&&{\cal H}_\perp=\displaystyle\frac{{\bf e}^{2/3}}{2T}Z_M{\cal M}^{MN}Z_N~~\nn\\
&&~~~{\cal M}^{MN}=(\nu^T)^{N}{}_B\hat{\delta}^{BA}\nu_A{}^M
~~,~~
\hat{\delta}^{AB}=\left(
{\renewcommand{\arraystretch}{1.4}
\begin{array}{ccc}
\delta^{ab}&0&\\
0&\delta_{[b_1}^{a_1}\delta_{b_2]}^{a_2}&0\\
0&0&1
\end{array}}
\right)\nn\\
&&~~~\nu_A{}^M=\left(
{\renewcommand{\arraystretch}{1.4}
\begin{array}{ccc}
{\bf e}^{-2/5}e_a{}^m&{\bf e}^{-2/5}C_{am_1m_2}^{[3]}&
\frac{1}{2}{\bf e}^{-2/5}{C}^{[3]}_{an_1n_2}\tilde{C}^{[3]n_1n_2}\\
0&{\bf e}^{-1/5}e_{[m_1}{}^{a_1}e_{m_2]}{}^{a_2}&
{\bf e}^{-1/5}\tilde{C}^{[3]a_1a_2}\\
0&0&{\bf e}
\end{array}}
\right)\nn\\
&&{C}^{[3]}_{am_1m_2}=e_a{}^mC^{[3]}_{mm_1m_2}
\label{nunu}\\&&
\tilde{C}^{[3]a_1a_2}=e_{m_1}{}^{a_1}e_{m_2}{}^{a_2}
\epsilon^{m_1m_2n_1n_2n_3}
C^{[3]}_{n_1n_2n_3}/3!
\nn\\
&&
{\bf e}=\det e_m{}^a=\sqrt{\det G_{mn}},\nn
\eea 
where  a normalization $\det \nu=1 $ is chosen.
The generalized metric ${\cal M}=\nu^T \nu$ is written in terms of
the metric and the gauge field, $(G_{mn}, ~C_{mnl}^{[3]})$, 
 which are 25 coset parameters 
of the coset SO(5,5)/SO(5)$\times$SO(5).
 Under the SO(5,5) transformations $Z_M$
and  $\nu_A{}^M$ are transformed as 
\bea
&Z_M\to(1+S)_M{}^NZ_N~,~
\nu_A{}^M \to (1+S_h)_A{}^B\nu_B{}^N (1+S)^{-1}{}_N{}^M~
&
~~~.
\eea
The transformation matrices $(1+S)\in$SO(5,5) and $(1+S_h)\in {\rm SO(5)\times SO(5)}$ are
 given in \bref{SO55S} and 
 \bref{SO5SO5S} where
the pullback parameter is determined  as
$\bar{\omega}_{ab}={\bf e}^{6/5}e_a{}^me_b{}^n\beta_{mn}$.
Although the SO(5,5) transformation of $e_m{}^a$ depends on SO(5) parameter $\omega$ as 
\bea
&&~~\delta e_m{}^a~=~\displaystyle\frac{3}{10}(-\hat{\alpha}+
\beta_{nl}\tilde{C}^{[3]nl})e_m{}^a
+\tilde{C}^{[3]al}\beta_{lm}-\omega_{m}{}^a+\alpha_m{}^a~~~,\nn
\eea
the transformations rules of SO(5) invariant gauge fields does not depend on it.
The obtained SO(5,5) transformation rules  of the metric and the 
gauge field are given as
\bea
&&\left\{
{\renewcommand{\arraystretch}{1.4}
\begin{array}{rcl}
\delta G_{mn}&=&\displaystyle\frac{3}{5}(-\hat{\alpha}
+\beta_{l_1l_2}\tilde{C}^{[3]l_1l_2})G_{mn}
+\beta_{l_1(m}G_{n)l_2}\tilde{C}^{[3]l_2l_1}
+\alpha_{(m}{}^lG_{n)l}\\
\delta C^{[3]}_{mnl}&=&
(-\hat{\alpha}+\displaystyle\frac{1}{2}\beta_{l_1l_2}\tilde{C}^{[3]l_1l_2})
C_{mnl}^{[3]}+\frac{1}{2}\alpha_{[m}{}^{l_1}C^{[3]}_{nl]l_1}
-{\bf e}^{12/5}\tilde{\tilde{\beta}}_{mnl}+\tilde{\gamma}_{mnl}
\end{array}}\right.\\
&&~~~~~\tilde{\tilde{\beta}}_{mnl}~=~
\frac{1}{2}\epsilon_{mnll_1l_2}G^{l_1n_1}G^{l_2n_2}\beta_{n_1n_2}~~~.
\nn
\eea
The transformation rule of $C^{[3]}$ is the fractional linear transformation 
as expected.
\par\vskip 6mm

\section{Extended space}

An extended space with manifest SO(5,5) duality symmetry 
is proposed  by introducing 
coordinates
$X^M=(x^m,~y_{mn},~y)$. They are subject to subsidiary conditions on functions $f(X^M)$
\bea
&
\partial_M\tilde{\rho}^{MN}_{[M2]}\partial_N
=\partial_M\tilde{\rho}^{MN}_{[M5]}\partial_N
=0&\nn~~~,
\eea
where 
$\partial_M=\frac{\partial}{\partial X^M}$ and 
$\tilde{\rho}^{MN}_{[M2]}$ and  $\tilde{\rho}^{MN}_{[M5]}$ are given  in \bref{diffeoZZ}. 
In components  they are  
\bea
&\displaystyle\frac{\partial}{\partial x^m}\frac{\partial}{\partial y_{mn}}=
\frac{\partial}{\partial x^m}
\frac{\partial}{\partial y}+\frac{1}{8}
\epsilon_{mm_1m_2m_3m_4}\frac{\partial}{\partial y_{m_1m_2}}
\frac{\partial}{\partial y_{m_3m_4}}
=0\nn~~,~~m=1,\cdots,5~~.
&
\eea
The extended space has manifest SO(5,5) symmetry, 
where SO(5,5) spinor coordinates are transformed as
$\delta X^M=-X^NS_N{}^M$ and $\delta \partial_M=S_M{}^N\partial_N$
preserving the canonical bracket 
$~\left\{\partial_M,X^N\right\}=\delta_M^N~$.
Extended coordinates are transformed  as
\bea
\left\{\begin{array}{ccl}
\delta x^m&=&\displaystyle
\frac{\hat{\alpha}}{2}x^m
-x^n\alpha_n{}^m-\frac{1}{2}y_{n_1n_2}
\tilde{\beta}^{n_1n_2m}\\
\delta y_{m_1m_2}&=&
x^n\displaystyle\tilde{\gamma}_{nm_1m_2}
-\frac{\hat{\alpha}}{2}y_{m_1m_2}
-y_{l[m_1}\alpha_{m_2]}{}^{l}
-y\beta_{m_1m_2}\\ 
\delta y&=&\displaystyle\frac{1}{2}
y_{n_1n_2}\gamma^{n_1n_2}-
y\frac{\hat{\alpha}}{2}
\end{array}\right.~~~.
\eea
On the other hand under SO(5)$\times$SO(5) $\ni$ $h$
satisfies $h^T\hat{\delta} h=\hat{\delta}$. 
Under the SO(5)$\times$SO(5) $\ni$ $h$
with  ``flat coordinates",
$x^a=e_m{}^a x^m$ ,~$y_{ab}=e_a{}^{m_1}e_b^{m_2}y_{m_1m_2}$~and 
$y$, are transformed as
\bea
&&\left\{\begin{array}{ccl}
\delta x^a&=&-\displaystyle\frac{1}{2}
x^b\omega_+{}_{;b}{}^a-\frac{1}{4}y_{b_1b_2}
\tilde{\omega}_+{}^{b_1b_2a}\\
\delta y_{a_1a_2}&=&\displaystyle\frac{1}{2}
x^b\displaystyle
\tilde{\omega}_+{}_{;ba_1a_2}-\displaystyle\frac{1}{2}
y_{b[a_1|}\omega_+{}_{|a_2]}{}^{b}-\displaystyle\frac{1}{2}
y\omega_+{}_{;a_1a_2}\\ 
\delta y&=&\displaystyle\frac{1}{4}
y_{b_1b_2}\omega_+{}^{b_1b_2}
\end{array}\right.\nn\\&&\label{SO5+}\\
&&\left\{\begin{array}{ccl}
\delta x^a&=&-\displaystyle\displaystyle\frac{1}{2}
x^b\omega_-{}_{;b}{}^a+\displaystyle\frac{1}{4}
y_{b_1b_2}
\tilde{\omega}_-{}^{b_1b_2a}\\
\delta y_{a_1a_2}&=&-\displaystyle\frac{1}{2}
x^b\displaystyle
\tilde{\omega}_-{}_{;ba_1a_2}-\displaystyle\frac{1}{2}
y_{b[a_1|}\omega_-{}_{|a_2]}{}^{b}
+\displaystyle\frac{1}{2}
y\omega_-{}_{;a_1a_2}\\ 
\delta y&=&-\displaystyle\frac{1}{4}
y_{b_1b_2}\omega_-{}^{b_1b_2}
\end{array}\right.~~~.\nn
\eea 

A SO(5,5) covariant  C-bracket in this extended space is  
proposed as
\bea
&\Lambda=\Lambda^M(X)Z_M
=\Lambda^M(x^m,y_{m_1m_2},y)Z_M&
\nn\\
&\left(\Bigl[\Lambda_1,\Lambda_2\Bigr]_C{}\right)^M=\Lambda^N_{[1}\partial_N\Lambda_{2]}^M
+\Lambda_1^N f_{NL}^{M;K}\partial_K\Lambda_2^L&
\eea 
 where $f_{NL}^{M;K}$  is SO(5,5) covariantized 
 version of the  symmetric structure constant $f_{NL}^{M;i}$  in \bref{rhof}.  
This bracket is reduced to the Courant bracket for the M5 brane 
obtained in \bref{CM5K} 
with the choice $\partial/\partial y_{m_1m_2}=\partial/\partial y=0$ and $K=-1$.

\par\vskip 6mm

\section{Summary and discussion}

We have  presented a M5 algebra from 
canonical analysis of a M5-brane in the supergravity background.
The M5 algebra is closed by the Gauss law constraint
which is the first class part of the selfduality condition.
The second class constraints of the 
selfduality condition are  treated by the Dirac bracket.
We have derived a series of  Courant brackets on the generalized geometry
 $T\oplus \Lambda^2T^\ast \oplus \Lambda^5T^\ast$
from the M5 algebra.
The Dirac bracket between selfdual gauge fields gives
 a $C^{[3]}$-twisted term in the Courant bracket.
By using it 
generalized diffeomorphism transformations including 
gauge transformations for $C^{[3]}$ and $C^{[6]}$ are derived.

The M-theory compactified on five dimensional torus  has
SO(5,5) duality symmetry, where 
the M5 algebra basis is in 16 dimensional SO(5,5) spinor representation.
The worldvolume diffeomorphism constraints 
are written in bilinear forms  in the M5 algebra basis, and 
they form a  10 dimensional SO(5,5) vector.
The generalized metric of ${\cal H}_\perp$ contains the metric and three form gauge field,
$G_{mn}$ and $C^{[3]}_{mnl}$, which are $15+10=25$ parameters of the coset 
SO(5,5)/[SO(5)$\times$SO(5)].

We have also proposed an extended space with manifest SO(5,5) duality symmetry 
with section conditions determined from the M5 worldvolume diffeomorphism constraints.
The SO(5,5) covariant C-bracket is also written down.

\par \medskip 
So far we have derived Courant brackets for M-theory in $d=4,5$ by
M2 and M5 algebra, while exceptional generalized geometry in the  brane algebra approach  
has not been discussed yet. 
We have analyzed only bosonic sector of supergravity of the M-theory. 
Inclusion of  fermions is an interesting issue for which
there are several  prior researches e.g. \cite{Hohm:2011zr}.
U-duality is raised by supersymmetry while supersymmetric probe branes
have $\kappa$-symmetry in addition to the worldvolume diffeomorphism constraints.
It is interesting to clarify the role of these constraints
in the  generalized geometry.  These issues are left for future investigations.

\section*{Acknowledgements}
M.H. would like to thank Yutaka Matsuo, Warren Siegel and Maxim Zabzine 
for fruitful discussions especially on M5 and selfdual gauge field. 
She also thanks  to Satoshi Iso and Takeshi Morita for valuable discussions.
 The work of M.H. is supported  by Grant-in-Aid for Scientific Research (C) No. 24540284 from The Ministry of Education, Culture, Sports, Science and Technology of Japan.

\end{document}